\documentclass[12pt]{iopart}
\usepackage{iopams}  
\begin{document}

\title[Site-bond representation and self-duality for TPCA]{Site-Bond Representation and Self-Duality for Totalistic Probabilistic Cellular Automata in One Dimension}

\author{N. Konno\footnote[1]{To
whom correspondence should be addressed (norio@mathlab.sci.ynu.ac.jp)}, X. Ma}
\address{Department of Applied Mathematics, Yokohama National University, Hodogaya-ku, Yokohama, 240-8501, Japan}

\begin{abstract}

We study the one-dimensional two-state totalistic probabilistic
 cellular automata (TPCA) having an absorbing state with long-range 
 interactions, which can be considered as a 
 natural extension of the Domany-Kinzel model. We establish the conditions for existence of a site-bond representation and self-dual property. Moreover we present an expression of a set-to-set connectedness between two sets, a matrix expression for a condition of the self-duality, and a convergence theorem for the TPCA. 

\end{abstract}

%\begin{keywords}
%percolation problems(theory), 
%phase transitions into absorbing states(theory)}
%\end{keywords}

%\ArXiv ePrint: cond-mat/0407657



%Comment out if separate title page not required
\maketitle

\section{Introduction}
Probabilistic versions of cellular automata can be considered as discrete-time Markov processes with parallel updating, which are useful in a large number of scientific areas \cite{Wolfram}. As a special case, the Domany-Kinzel (DK) model introduced and studied by \cite{DK} is one of the simplest examples of the probabilistic cellular automata. The model is defined on a lattice, and the two different states for a site can be said as empty or occupied. The state
 of a given site depends on the number of the occupied 
 states in its two nearest neighbors at the previous time step, where 
 this property can be expressed with the {\it totalistic} rule \cite{Wolfram,Bagnoli,BR}. In the DK model, the transition probability from an empty neighborhood to an occupied state is zero, so the empty configuration is an absorbing state. The totalistic probabilistic cellular automaton (TPCA) is a natural extension of the DK model, where we extend the number of neighbors to a finite integer $N (\ge 2)$. Remark that $N=2$ case becomes the DK model. Concerning the DK model, various results are known (see our recent papers \cite{kkt,kk2000,kkt2002,kkst} and references therein). As for $N=3$ case with two absorbing states, a rich phase diagram is reporeted by using mean-field approximations and numerical simulations (see Refs. \cite{BBR,ADM}, for examples). However there are few rigorous results on the TPCA for a general $N \ge 3$. In this situation, the purpose of this paper is to give some rigorous results for a two-state $N$-neighbor TPCA with an absorbing state. More precisely, we reveal a relation between the site-bond representation and the self-duality for the TPCA. Furthermore we present an expression of a set-to-set connectedness between two sets, a matrix expression for a condition of the self-duality, and a convergence theorem for the TPCA. It is known that the self-duality is a very useful technique in the study of stochastic interacting models. Because problems in uncountable state space (typically configurations of $\{0,1\}^{\bf Z}$) can be reformulated as problems in countable state space (typically finite subsets of ${\bf Z}$). For some applications of the self-duality on discrete-time case (e.g., oriented bond percolation model) and continuous-time case (e.g., contact process), see Refs. \cite{Liggett,Durrett,Konnobook,LiggettNew}. So our results here might be useful in obtaining rigorous results on the phase diagram and critical properties of such models. 

We have organized the present paper as follows: Section 2 is devoted to the definition of our TPCA. In Section 3, we explain a site-bond representation and a self-duality for the TPCA. Moreover an expression of a set-to-set connectedness for the TPCA is given. Section 4 treats a matrix expression for the self-duality of the model. Section 5 contains a convergence theorem. Finally, concluding remarks and discussions are presented in Section 6. 

\section{Definition of the TPCA}
First we give the definition of the DK model ($N=2$ case). Let $ \xi_n ^A \subset {\bf Z}$ be the state of the process with parameters $(p_1,p_2) \in [0,1]^2 $ at time $n$ starting from a set $A \subset 2{\bf Z}$. Its evolution is described by
\[
P(x \in \xi_{n+1} ^A | \xi_n ^A) = f(\xi_n^A (x-1) + \xi_n^A (x+1)),
\]
and given $ \xi_n^A,$ the events $\{x \in \xi_{n+1}^A \}$ are independent, where $P(B|C)$ is the conditional probability that $B$ occurs given that $C$ occurs, and 
\[
f(0)= 0, \quad  f(1)= p_1, \quad f(2)= p_2.
\]
If we write $ \xi(x,n)=1 $ for $ x\in\xi_n^A $ and $ \xi(x,n)=0 $ otherwise, each realization of the process is identified with a configuration $ \xi \in
\{ 0,1 \} ^S$ with $S= \{ s=(x,n) \in {\bf Z} \times {\bf Z}_+:x+n= $ even $\}$, where ${\bf Z}_+= \{ 0,1,2,\ldots \}$. The model with $(p_1,p_2)=(1,0)$ becomes Wolfram's rule 90. For more detailed information, see Section 5 in \cite{Durrett}.

\par
From now on, we introduce a long-range TPCA. In order to clarify the definition, we consider two cases of $N=$ even and $N=$ odd respectively, where $N$ is the number of the neighborhood. 
\par
(i) $N=2L \> (L=1,2, \ldots)$ case. The space of sites is denoted by 
$ S_0= \{s=(x,n) \in {\bf Z} \times {\bf Z}_+ :x+n=$ even $\}$ and the space 
of bonds is
\begin{eqnarray*}
B_0 = \{ ((x,n+1),(x-2L+1,n)), \ldots, ((x,n+1),(x-1,n)), 
((x,n+1),\\ 
(x+1,n)),\ldots , 
((x,n+1),(x+2L-1,n)) : (x,n+1) \in S_0 \}.
\end{eqnarray*}
For any initial set $A \subset 2{\bf Z}$, the process $\xi_n ^A $
satisfies
\[
P(x \in \xi_{n+1} ^A | \xi_n ^A) =
f(\sum_{k=-(L-1)}^{L} \xi_n ^A (x-2k+1)),
\]                
and given $\xi_{n} ^A$, the events 
$\{x \in \xi_{n+1} ^A \}$ are independent, where       
\[
f(0)=p_0=0, f(1)=p_1, \ldots, f(2L)=p_{2L},
\]
with $ p_1,p_2,\ldots,p_{2L} \in [0,1] $. This process is 
considered on the space $ S_0 $.
The case $L=1$ is equivalent to the DK model.
\par
(ii) $N=2L+1 \> (L=1,2, \ldots)$ case. The space of sites is denoted by 
\[
S_1= \{ s=(x,n) \in {\bf Z} \times {\bf Z}_+ \},
\]
and the space of bonds is
\begin{eqnarray*}
&& B_1=\{((x,n+1),(x-L,n)), \ldots ,((x,n+1),(x-1,n)), ((x,n+1),
\\ 
&& (x,n)), 
 ((x,n+1),(x+1,n)), \ldots , ((x,n+1),(x+L,n)) : 
 (x,n+1) \in S_1 \}.
\end{eqnarray*}
For any initial set $A \subset {\bf Z}$, the process $\xi_n ^A$
satisfies
\[
P(x \in \xi_{n+1} ^A | \xi_n ^A) =
  f( \sum_{k=-L}^{L} \xi_n ^A (x+k)),
\] 
and given $\xi_{n} ^A$, the events $\{x \in \xi_{n+1} ^A \}$ 
are independent, where
\[       
f(0)=p_0=0, f(1)=p_1, \ldots, f(2L+1)=p_{2L+1},
\]
with $ p_1,p_2, \ldots, p_{2L+1} \in [0,1] $. 
This process is considered on the space $S_1.$ 
\par
If $p_0 = 0 \le p_1 \le p_2 \le \ldots \le p_{N-1} \le p_N$, then the $N$-neighbor TPCA is said to be {\it attractive}. If not, it is said to be {\it non-attractive}. The attractiveness means that having more particles at one time implies there will be more particles at the next time(s). 

\par
In general, few rigorous results on the non-attractive model are known compared with the attractive model. Much more informations on the results for the non-attractive $N=2$ case are shown for instance in \cite{kkt2002}. So studying a general model including the non-attractive case, such as is the one in this paper, is very important.

\section{Site-bond representation and self-duality}
From now on we consider only $(p_1,p_2) \in D_{\ast}$ case, where 
 the subset of parameter space $D_{\ast}$ is defined in the following: 
\[
D_{\ast} = \{ (p_1,p_2) :  0 < p_1 \le 1, \> 0 < p_2 \le 1 \> \mbox{and} 
\> p_2< 2p_1 \}.
\]
A reason for introducing the set is that for any point outside of 
 $D_{\ast}$, it will be shown that there exists no site-bond representation for the TPCA (see Proposition 1 below). Suppose that $ p_1, p_2 \in D_{\ast}$ are given. We put 
$\alpha=p_1^2/(2p_1-p_2), \> \beta =2- p_2/p_1$. Let $ (S,B)=(S_0,B_0)$ or $(S_1,B_1)$ 
and ${\bf R}$ denote the set of real numbers. If there exist $\alpha, \beta \in {\bf R}$ such that $p_n \in [0,1]$ is given by
\begin{eqnarray}
p_n = \alpha [1-(1-\beta)^n], 
%\label{eq:sbcon}
\label{eqn:one}
\end{eqnarray}
for $n=3,4, \ldots, N$, then we call ``an $N$-neighbor TPCA with $\{p_n : 0 \le n \le N \}$ has a site-bond representation with $\alpha$ and $\beta$''. In fact, when $\alpha, \beta \in (0,1]$, the TPCA can be considered as an oriented mixed site-bond percolation with a long range interaction in the following way. On the space $(S,B)$ we define 
\[
X(S)=\{0,1\}^S,\quad X(B)=\{0,1\}^B,\quad X=X(S) \times X(B).
\]
For given $ \zeta=(\zeta_1,\zeta_2) \in X $, we say
 that $s=(y, n+k) \in S $ can be reached from 
$ s'=(x,n) \in S $ and write $ s'\rightarrow s $,
 if there exists a sequence $ s_0, s_1, s_2, \ldots, s_k $ of 
 members of $S$ such that $ s'= s_0, s= s_k $, and 
\[
\zeta_1(s_i)=1,i=0,1, \ldots, k; \> 
\zeta_2(( s_i,s_{i+1}))=1,i=0,1, \ldots, k-1. 
\]
we also say that $ G \subset S $ can be reached from 
$ G'\subset S $ and write $ G'\rightarrow G $,
if there exist $ s \in G $ and $ s'\in G' $ 
such that $ s'\rightarrow s $.
\par
We introduce the signed measure $m$ on $X$ defined by 
\[         
m(\Lambda)=\alpha^{k_1} 
(1-\alpha)^{j_1}\beta^{k_2}(1-\beta)^{j_2},            
\]
for any cylinder set 
\begin{center} 
$                
 \Lambda=\{(\zeta_1,\zeta_2) \in X $: 
    $ \zeta_1(s_i)=1,i=1,2,\ldots,k_1 $;
    $ \zeta_1(s_i')=0,i=1,2,\ldots,j_1 $;\\
    $ \zeta_2 (b_i)=1,i=1,2,\ldots,k_2 $; 
    $ \zeta_2 (b_i')=0,i=1,2,\ldots,j_2 \}, $ 
\end{center}
where
$ s_1,s_2,\ldots, s_k, s_1', s_2' $,
$ \ldots, s_j' $ are distinct elements of $S$, 
and $ b_1,b_2 $,$ \ldots ,b_k $,$ b_1', b_2' $,$ \ldots,
 b_j' $ are distinct elements of 
$B$ and $ \alpha=p_1^2/(2p_1-p_2), \> \beta =2- p_2/p_1 $.
\par
We should remark the next three facts. If $ p_2 < 2p_1 $ and $ p_2 > 2p_1 - p_1^2 $, then $\alpha >1$ and $ \beta \in (0,1)$. If $p_2 \le 2p_1 - p_1^2 $ and $ p_2 \ge p_1 $, then $\alpha, \beta \in (0,1] $. Moreover, 
if $ p_2 < p_1 $, then $ \alpha \in (0,1) $ and $ \beta \in (1,2]$.
 From the above observation, we see that the measure
 is not probability measure in the first and third cases,
  since $ \alpha >1 $ and $ 1-\beta < 0 $ respectively. See Refs. \cite{kkt} and \cite{kkt2002} for more details on the site-bond representation. 

It is noted that if an $N$-neighbor TPCA has a site-bond representation with $\alpha$ and $\beta$, then $\alpha=p_1^2/(2p_1-p_2), \> \beta =2- p_2/p_1$, since Eq. (\ref{eqn:one}) gives
\[
p_1 = \alpha \beta, \qquad p_2 = \alpha [1 - (1- \beta)^2].
\]
As we will see in the proof of Proposition 1, the following result can be obtained: an $N$-neighbor TPCA having a site-bond representation with $\alpha$ and $\beta$ is attractive if and only if $p_1 \le p_2$. In other words, an $N$-neighbor TPCA having a site-bond representation with $\alpha$ and $\beta$ is non-attractive if and only if $p_1 > p_2$. Then we have
\\
\\
{\bf Proposition 1} 
We consider $N$-neighbor TPCA with $\{p_n : 0 \le n \le N \}$. Assume $(p_1,p_2) \in D_{\ast}$. Let $\alpha = p_1^2/(2p_1-p_2)$ and $\beta = (2p_1-p_2)/p_1$.
(i) $p_1 > p_2$ (non-attractive case) and $p_n $ $= \alpha [1 - (1-\beta)^n] \> (0 \le n \le N)$, then $p_n \in [0,1] \> (0 \le n \le N)$. (ii) $p_1 \le p_2$ (attractive case), $p_n $ $= \alpha [1 - (1-\beta)^n] \> (0 \le n \le N)$, and $p_N \le 1$, then $p_n \in [0,1] \> (0 \le n \le N)$.
%\end{pro}
\\
\\
{\bf Proof}. First we consider a relation among $p_0 = 0, p_1, \ldots, p_N$. By using $p_n $ $= \alpha [1 - (1-\beta)^n]$ for $n=0,1, \ldots , N$, we see 
\begin{eqnarray}
&& 
p_n - p_{n-1} = \alpha \beta (1- \beta )^{n-1}, 
\label{eqn:two}
\\
&&         
p_{n}-p_{n-2}
= \alpha (1- \beta )^{n-2} \{ 1-(1- \beta )^2 \}.  
\label{eqn:four}
\end{eqnarray}
Recall that
\begin{eqnarray}
1- \beta =(p_2-p_1)/p_1, \qquad \alpha \beta =p_1. 
\label{eqn:six}
\end{eqnarray}
Now we consider $p_1 > p_2$ (non-attractive) case. In this case, 
Eq. (\ref{eqn:six}) gives 
\begin{eqnarray}
-1 < 1-\beta < 0, \qquad 0 < \alpha \beta \le 1, \qquad 0< \alpha < 1. 
\label{eqn:seven}
\end{eqnarray}
From Eqs. (\ref{eqn:two}), (\ref{eqn:four}), and (\ref{eqn:seven}), we see  
\begin{eqnarray*}
p_0=0 < p_2 < p_4 < p_6 < \ldots < p_{2n} < p_{2n-1}
   < \ldots < p_5 < p_3 < p_1 \le 1.
\end{eqnarray*}
Thus we obtain $p_n \in [0,1]$ for any $n =0,1, \ldots, N$. Next we consider $p_1 \le p_2$ (attractive) case. It is easily checked that $p_n \le p_{n+1}$ for any $n=0,1, \ldots, N-1$ by using Eq. (\ref{eqn:two}) and $0 \le 1- \beta < 1$. So we have 
\begin{eqnarray*}
p_0 = 0 < p_1 \le p_2 \le p_3 \le \ldots \le p_{N-1} \le p_N. 
\end{eqnarray*}
Therefore a necessary and sufficient condition for the site-bond representation in this case is $p_N = \alpha [1 - (1- \beta)^N] \le 1$. The proof of Proposition 1 is complete.  
\par
\
\par\noindent
Remark that $p_n = \alpha [1 - (1- \beta)^n] \to \alpha$ as $n \to \infty,$ since $|1- \beta|<1$. Noting that $\alpha \le 1$ if and only if $p_2 \le 2p_1-p_1^2,$ we conclude that a sufficient (and not necessary) condition for a site-bond representation of an $N$-neighbor TPCA is $p_2 \le 2p_1-p_1^2$. Furthermore Eq. (\ref{eqn:two}) implies
%\begin{cor}
\\
\\
{\bf Corollary 2} 
If an $N$-neighbor TPCA with $\{ p_n : 0 \le n \le N\}$ and $N \ge 3$ has the site-bond representation, then $p_{n}$ can be expanded
with $p_1$ and $p_2 $ as follows:
\begin{eqnarray*}                    
p_n = 
{(p_2-p_1)^{n-1} \over p_1^{n-2}}+	   
{(p_2-p_1)^{n-2} \over p_1^{n-3}}+ \cdots +                
{(p_2-p_1)^2 \over p_1} + p_2,    
\label{eqn:eight}           
\end{eqnarray*}
for any $n = 3,4, \ldots , N.$
\\
%\end{cor}
\par                 
Next we define a set-to-set connectedness for the TPCA from a set $A$ to a set $B$ by
\[
\sigma (A,B) = \lim_{n \to \infty} P(\xi_{n}^A \cap B \neq \emptyset),
\]
if the right-hand side exists. As in a similar argument of $N=2$ case (the DK model) \cite{kkt}, we can easily extend the case to a general $N$-neighbor TPCA: 
%\begin{pro}
\\
\\
{\bf Proposition 3} 
Let $(p_1,p_2) \in D_{\ast}$. We assume that an $N$-neighbor TPCA with $\{ p_n : 0 \le n \le N\}$ has a site-bond representation with $\alpha$ and $\beta$, where $\alpha = p_1^2/(2p_1-p_2)$ and $\beta = (2p_1-p_2)/p_1$. Then for any $A$ 
with $|A|<\infty $ , we have
\[
\sigma (2 {\bf Z}, A) 
=
\sum_{D \subset A,D \neq \emptyset}
 \alpha^{|D|}(1-\alpha)^{|A \backslash D|}\sigma(D,2{\bf Z}),
\]
when $ N=2L $, and
\[
\sigma ({\bf Z},A) 
= \sum_{D \subset A,D \neq \emptyset} 
\alpha^{|D|}(1-\alpha)^{|A \backslash D|}\sigma(D,{\bf Z}),
\] 
when $ N=2L+1.$
%\end{pro}
\par
\
\par
From now on, we consider a self-duality of the TPCA. An $N$-neighbor TPCA $\xi_n$ is said to be {\it self-dual} with a {\it self-duality parameter} $x$ if
\[
E \left( x^{|\xi^{A} _n \cap B |} \right)
= E \left( x^{|\xi^{B} _{n} \cap A |} \right) \qquad (n=0,1,\ldots )
\]
holds for any $A, B \subset {\bf Z}$ with $|A| < \infty$ or $|B| < \infty$. The above equation is called a {\it self-duality equation}. Then we have
%\begin{thm}
\\
\\
{\bf Theorem 4} 
Let $(p_1,p_2) \in D_{\ast}$ with $\alpha [1 - (1- \beta)^N] \le 1$. An $N$-neighbor TPCA with $\{ p_n : 0 \le n \le N\}$ is self-dual with a self-duality parameter $(\alpha -1)/\alpha$ if and only if this model has a site-bond representation with $\alpha$ and $\beta$, where $\alpha = p_1^2/(2p_1-p_2)$ and $\beta = (2p_1-p_2)/p_1$.
%\end{thm}
\\
{\bf Proof.} Theorem 1 in \cite{Konnoa} implies that an $N$-neighbor TPCA with transition probabilities $\{ p_n : 0 \le n \le N\}$ is self-dual with a self-duality parameter $x$ is equivalent to 
\begin{eqnarray}
(p_i x + q_i)^j = (p_j x + q_j)^i,
\label{eqn:symm}
\end{eqnarray}
for $0 \le i,j \le N$, where $ q_i = 1- p_i$. It can be confirmed that $x=(\alpha -1)/ \alpha$ with $\alpha=p_1^2/(2p_1-p_2)$ satisfies Eq. (\ref{eqn:symm}). Moreover we see that Eq. (\ref{eqn:symm}) if and only if 
\begin{eqnarray}
(p_1 x + q_1)^n = p_n x + q_n, 
\label{eqn:iter}
\end{eqnarray}
for $0 \le n \le N$. In fact if we take $i=1$ and $j=n$, Eq. (\ref{eqn:symm}) becomes 
Eq. (\ref{eqn:iter}). Conversely we have
\[
(p_i x + q_i)^j = (p_1 x + q_1)^{ij}= (p_j x + q_j)^i. 
\] 
Therefore  we put $y=x-1 $. Then 
Eq. (\ref{eqn:iter}) becomes $p_n y + 1 = (p_1 y +1)^n$. So we see that 
\begin{eqnarray*}
p_n &=& \sum_{r=1}^{n}
     {n \choose r} 
          y^{r-1}p_1^r \\
&=& \sum _{r=1}^{n}{n \choose r}
    \left( {p_2-2p_1 \over p_1^2} \right)^{r-1} p_1^r \\                        
&=& \left( {p_2-2p_1 \over p_1^2} \right)^{-1} \sum_{r=1}^{n}{n \choose r}
\left( {p_2-2p_1 \over p_1} \right)^r \\  
&=& \alpha  \sum_{r=0}^{n} 
{n \choose r} \left(1 - {p_2-2p_1 \over p_1} \right)^r  \\  
&=& \alpha [1-(1+(p_2-2p_1)/p_1)^n ] \\
&=& \alpha [ 1-(1-\beta)^n ].
\end{eqnarray*}
So the proof of Theorem 4 is complete.
\par
\
\par
From this theorem it can be seen that two conditions for the self-duality and for the site-bond representation are equivalent under an assumption that $(p_1,p_2) \in D_{\ast}$ with $p_N = \alpha [1 - (1- \beta)^N] \le 1$.

\section{Matrix expression}
In this section we consider a criterion for the self-duality based on a matrix expression. Let 
\begin{eqnarray*}
X(x) = 
\left[
\begin{array}{cc}
1 & 1 \\
1 & x 
\end{array}
\right],
\end{eqnarray*}
and 
\[
X_N (x) = X (x)^{\otimes N},
\]
where $\otimes$ indicates the tensor product. For example, in the case of $N=2$, we have
\begin{eqnarray*}
X_2 (x) = X (x) \otimes X (x) = 
\left[
\begin{array}{cccc}
1 & 1 & 1 & 1 \\
1 & x & 1 & x \\
1 & 1 & x & x \\
1 & x & x & x^2 
\end{array}
\right].
\end{eqnarray*}
For fixed $i,j \in {1,2, \ldots , 2^{N}}$, the values of 
 $i_1, i_2, \ldots , i_N$ and $j_1, j_2, \ldots , j_N \in \{0,1\}$ are 
 defined by the binary expansion of $i$ and $j$ in the following: 
\begin{eqnarray*}
&& 
i-1 = i_1 2^{N-1} + i_2 2^{N-2} + \cdots + i_{N-1} 2^1 + i_N 2^0,
\\
&&
j-1 = j_1 2^{N-1} + j_2 2^{N-2} + \cdots + j_{N-1} 2^1 + j_N 2^0,
\end{eqnarray*}
and $I=i_1+i_2+ \cdots +i_N$. Moreover we introduce $2^N \times 2^N$ matrix $P_N$ whose $(i,j)$ element is defined by
\begin{eqnarray*}
p_I ^{(j_1)} p_I ^{(j_2)} \cdots p_I ^{(j_N)}, 
\end{eqnarray*}
where $p_I ^{(1)} = p_I, \> p_I ^{(0)} =1 - p_I = q_I$. 
 For $N=2$ case, the above definition gives
\begin{eqnarray*}
P_2= 
\left[
\begin{array}{cccc}
1 & 0 & 0 & 0 \\
q_1^2 & p_1 q_1 & p_1 q_1 & p_1^2 \\
q_1^2 & p_1 q_1 & p_1 q_1 & p_1^2 \\
q_2^2 & p_2 q_2 & p_2 q_2 & p_2^2 
\end{array}
\right],
\end{eqnarray*}
since $p_0 =0, \> q_0=1$. Then we have the next result: 
%\begin{thm}
\\
\\
{\bf Theorem 5} 
A necessary and sufficient condition of the self-duality for an $N$-neighbor TPCA with $\{ p_n : 0 \le n \le N\}$ with a self-duality parameter $x$ is that $x$ satisfies
\begin{eqnarray}
P_N X_N (x) = {}^t (P_N X_N (x)),
\label{eqn:eleven}
\end{eqnarray}
where $t$ means the transposed operater. 
%\end{thm}
\\
{\bf Proof}. Let $c^{(N)} _{ik} (x)$ denote $(i,k)$ element of $P_N X_N (x)$. First $i_1, i_2, \ldots , i_N$ and $k_1, k_2, \ldots , k_N$ are defined by $i-1 = i_1 2^{N-1} + i_2 2^{N-2} + \cdots + i_{N-1} 2^1 + i_N 2^0$ and $k-1 = k_1 2^{N-1} + k_2 2^{N-2} + \cdots + k_{N-1} 2^1 + k_N 2^0$, respectively. We put $I=i_1+i_2+ \cdots +i_N$ and $K=k_1+k_2+ \cdots +k_N$. Moreover $\{ k_1, k_2, \ldots , k_N \}$ is divided into two sets such as $\{ k_{m_1}, k_{m_2}, \ldots k_{m_K} : k_u =1$ for any $u \in \{ m_1, m_2, \ldots, m_K \} \}$ and $\{ k_{l_1}, k_{l_2}, \ldots k_{l_{N-K}} : k_v =0$ for any $v \in \{l_1, l_2, \ldots, l_{N-K} \} \}$. Then we see that
\begin{eqnarray*}
c^{(N)} _{ik} (x) 
&=&
\sum_{j_{m_1}, \ldots, j_{m_K} \in \{0,1\}} 
\sum_{j_{l_1}, \ldots, j_{l_{N-K}} \in \{0,1\}}
p_I ^{(j_{m_1})} \cdots p_I ^{(j_{m_K})} \times
p_I ^{(j_{l_1})} \cdots p_I ^{(j_{l_{N-K}})}
\\
&&
\qquad \qquad \qquad \qquad \qquad 
\times x_{(j_{m_1})} ^{k_{m_1}} \cdots x_{(j_{m_K})} ^{k_{m_K}}
\times 
x_{(j_{l_1})} ^{k_{l_1}} \cdots x_{(j_{l_{N-K}})} ^{k_{l_{N-K}}}
\\
&=&
\sum_{j_{m_1}, \ldots, j_{m_K} \in \{0,1\}}
p_I ^{(j_{m_1})} \cdots p_I ^{(j_{m_K})}
x_{(j_{m_1})} ^{k_{m_1}} \cdots x_{(j_{m_K})} ^{k_{m_K}}
\\
&=&
 (p_I ^{(1)} x_{(1)} + p_I ^{(0)} x_{(0)} )^K
\\
&=&
(p_I x + q_I)^K.
\end{eqnarray*}
where $x_{(1)} =x$ and $x_{(0)}=1$. In a similar way, we have
\[
c^{(N)} _{ki} (x) = (p_K x + q_K)^I.
\]
So we have
\begin{eqnarray*}
P_N X_N (x) = {}^t (P_N X_N (x)),
\end{eqnarray*}
if and only if
\begin{eqnarray*}
c^{(N)} _{ik} (x) = c^{(N)} _{ki} (x), 
\end{eqnarray*}
for $1 \le i,k \le 2^N$. Forthermore it is shown that the last equation is equivalent to
\begin{eqnarray*}
(p_I x + q_I)^K = (p_K x + q_K)^I,  
\label{eq:sym}
\end{eqnarray*}
for $1 \le I,K \le N.$ Combaining the above result with Theorem 1 in \cite{Konnoa}, we obtain the desired conclusion.
\par
\
\par
For $N=2$ case (the DK model), Theorem 5 gives 
\[
(p_1 x + q_1)^2 = p_2 x + q_2.
\]
This result was shown in \cite{kkst,Konnoa,Konnob} by using different methods.

\section{Convergence theorem}
In order to obtain a convergence theorem for the TPCA with $\{ p_n : 0 \le n \le N\}$ having a site-bond representation with $\alpha = p_1^2/(2p_1-p_2)$ and $\beta = (2p_1-p_2)/p_1$, which corresponds to the result given by \cite{kkst}, a new process $\eta_n$ is introduced as follows. For simplicity, we first consider the DK model ($N=2$ case) with $(p_1,p_2) \in D_{\ast}$. Put $p_{\ast} = \max \{p_1, p_2\}.$ A new process defined below is called {\it $p_{\ast}$-DK-dual}. We can see the thinning-relationship by coupling the DK model and $p_{\ast}$-DK-dual. We split both the DK model and the $p_{\ast}$-DK-dual into two phases, and we will allow the first phase to occur at times $n+(1/2)$ for $n \in {\bf Z_{+}}.$

\begin{enumerate}
\item  Let $\mu $ be the distribution of the $p_{\ast}$-DK-dual at time $0.$

\item  At time $n=1/2,$ it undergoes a $p_{\ast}$-thinning. In general, for $p \in [0,1],$ the $p$-thinning of a set $A \subset {\bf Z}$ is the random subset of $A$ obtained by independently removing each element of $A$ with probability $1-p.$

\item  Start the DK model at time $n=0$ with the same
configuration as the $p_{\ast}$-DK-model (which is defined by $f(0)=0, \> f(1)=p_1/p_{\ast}, \> f(2)=p_2/p_{\ast}$) at time $n=1/2.$ 

\item  Couple the processes together until time $n_0-(1/2)$ for the
DK model, $n_0$ for the $p_{\ast}$-DK-model. This can be done because the
transitions for the DK model are the same as those for the
$p_{\ast}$-DK-model lagged by time unit $1/2.$

\item  Perform a $p_{\ast}$-thinning for the DK model at time $n_0.$
\end{enumerate}

\noindent The distribution of the DK model started and ended as a 
$p_{\ast}$-thinning of the $p_{\ast}$-DK-dual. As in the DK model, we can define a new process $\eta_n$ as {\it $p_{\ast}$-TPCA-dual} for the $N$-neighbor TPCA, where $p_{*}=\max\{p_{1},p_{2},\ldots,p_{N}\}$. Recall that the following duality holds for the TPCA $\xi^A _n$ and the $p_{\ast}$-TPCA-dual $\eta^A _n$ starting from $A,$ (see Theorem 2 (2) in \cite{Konnoa}). 
\\
\\
%\begin{thm}
{\bf Theorem 6 (\cite{Konnoa})} We assume an $N$-neighbor TPCA with $\{ p_n : 0 \le n \le N\}$ having a site-bond representation with $\alpha = p_1^2/(2p_1-p_2)$ and $\beta = (2p_1-p_2)/p_1$. Suppose $(p_1,p_2) \in D_{\ast}$. For any $A,B$ with $|A| < \infty$ or $|B| < \infty$, we have 
\[
E \left(\tilde{x}^{|\xi^{A} _n \cap B |} \right)
= E \left( \tilde{x}^{|\eta^{B} _{n} \cap A |} \right), 
\]
for any $n \ge 0$, if $\tilde{x}=1- (2p_1 - p_2)p_*/p_1^2$.
%\end{thm}
%and from the proof of Proposition 1 we know $p_*=p_1$, when $p_1 > p_2$.
\\
\\
Note that if $p_1 > p_2$ (non-attractive case for the TPCA with a site-bond representation), then $p_*=p_1$. By using the same argument for the DK model in Section 7 of \cite{kkst}, we can get the following convergence theorem for the TPCA: 
%\begin{thm}
\\
\\
{\bf Theorem 7} Assume $p_1 > p_2$ with $(p_1,p_2) \in D_{\ast}$. If the initial measure $\nu$ of the TPCA is a.s. (almost sure) infinite, then we have
\[
\xi_n ^{\nu} \to \mu_{\eta}, 
\]
as $n \to \infty$, where the limit measure is uniquely determined by 
$E ((-(p_1 - p_2)/p_1)^{|\mu_{\eta} \cap A|})= P(|\eta_{\infty}^A|=0)$ for any $A$ with $|A|< \infty.$
%\end{thm}

\section{Conclusions and discussions}
In this work we have presented rigorous results on the site-bond representation, the set-to-set connectedness, the self-duality, the matrix expression, and the convergence theorem for an $N$-neighbor TPCA with $\{p_n : 0 \le n \le N \},$ where $p_0 =0$ and $N \ge 2$. An interesting feature of our model is that the dominant parameters on some properties are only $p_1$ and $p_2$ among $\{p_n : 0 \le n \le N \}$, see Proposition 1 and Corollary 2, for examples.
\par
Arrowsmith and Essam \cite{Arrowsmith} gave an expansion formula for a point-to-point connectedness for the oriented mixed site-bond percolation, in which each term is characterized by a graph. Konno and Katori \cite{kk2000} extended this formula to $N=2$ case (the DK model). So it is shown that site-bond representation, self-duality, and the above graphical expansion formula hold in the DK model. Thus one of the interesting future problems is to extend the relation to a general $N$-neighbor TPCA considered here. 
\par
Finally we mention a relation between our discrete-time model and a continous-time one corresponding to it. In the continuous-time case, an infinitesimal generator for the model corresponds to $P_N$ in Theorem 5. In fact, Eq. (\ref{eqn:eleven}) in Theorem 5 corresponds to Eq. (17) in \cite{sl1995} for $N=2$ case (an extension of their result to an $N$-neighbor case can be easily obtained). Recent works on the duality for continuous-time interacting particle systems are presented in \cite{Bandt,Sudbury2000,LS,HV}.
\\
\\
\par\noindent
{\bf Acknowledgments.} We would like to thank Akinobu Shimizu for useful discussions. This work was partially financed by the Grant-in-Aid for Scientific Research (B) (No.12440024) of Japan Society of the Promotion of Science. 
\\
\\

\end{document}